\documentclass[modern]{aastex631}%trackchanges
\usepackage[normalem]{ulem}
\usepackage{textcomp, gensymb, amsmath}
\useunder{\uline}{\ul}{}

\submitjournal{ApJ}

\shorttitle{Short-term SEP Event Forecasting}
\shortauthors{Rotti et al.}
\begin{document}

\title{Precise and Accurate Short-term Forecasting of Solar Energetic Particle Events with Multivariate Time Series Classifiers}

\correspondingauthor{Sumanth A. Rotti}
\email{srotti@gsu.edu}

\author[0000-0003-1080-3424]{Sumanth A. Rotti}
\affiliation{Georgia State University \\
Department of Physics and Astronomy \\
Atlanta, GA, USA}

\author[0000-0002-9799-9265]{Berkay Aydin}
\affiliation{Georgia State University \\
Department of Computer Science \\
Atlanta, GA, USA}

\author[0000-0001-8078-6856]{Petrus C. Martens}
\affiliation{Georgia State University \\
Department of Physics and Astronomy \\
Atlanta, GA, USA}

\begin{abstract}
Solar energetic particle (SEP) events are one of the most crucial aspects of space weather that require continuous monitoring and forecasting using robust methods. We demonstrate a proof of concept of using a data-driven supervised classification framework on a multivariate time series data set covering solar cycles 22, 23, and 24. We implement ensemble modeling that merges the results from three proton channels (E$\geq$10 MeV, 50 MeV, and 100 MeV) and the long band X-ray flux (1–8Å) channel from the Geostationary Operational Environmental Satellite missions. Our task is binary classification, such that the aim of the model is to distinguish strong SEP events from nonevents. Here, strong SEP events are those crossing the Space Weather Prediction Center's ``S1'' threshold of solar radiation storm and proton fluxes below that are weak SEP events. In addition, we consider periods of non-occurrence of SEPs following a flare with magnitudes $\geq$C6.0 to maintain a natural imbalance of sample distribution. In our data set, there are 244 strong SEP events comprising the positive class. There are 189 weak events and 2,460 ``SEP-quiet'' periods for the negative class. We experiment with summary statistic classifier, one-nearest neighbor and supervised time series forest (STSF), and compare their performances to validate our methods for prediction windows from 5 min up to 60 min. We find STSF to perform better under all circumstances. For an optimal classification threshold of $\approx$0.3 and a 60 min prediction window, we obtain: TSS = 0.850, HSS = 0.878, GSS = 0.783. 
\end{abstract}


\section{Introduction} \label{sec:intro}
The activity of the Sun is considered the prime source of space weather (SWx) that constitutes different eruptive phenomena including solar flares (SFs) and coronal mass ejections (CMEs). SF is a sudden brightening in the solar atmosphere in a coronal soft X-ray source that is observed to have close relationships with CMEs (Feynman \& Hundhausen \citeyear{feynman1994coronal}) and other observable phenomena such as jets and filament eruptions (Schrijver \& Siscoe \citeyear{schrijver2010heliophysics}). CMEs are clouds of plasma formed in the lower corona that often move faster ($>$1000 km.s$^{-1}$) than the ambient solar wind (Low \citeyear{low1996solar}; Chen \citeyear{chen2011coronal}). Another manifestation of solar activity constituting the emission of energetic electrons, protons, and heavier ions from the Sun is called a solar energetic particle (SEP) event (Klein \& Dalla \citeyear{klein2017acceleration}). Typically, large flares and shock fronts of fast CMEs are known to accelerate SEPs and are often considered precursors or parent eruptions (Cane et al. \citeyear{1986cane}; Kahler \citeyear{1992kahler}; Gopalswamy et al. \citeyear{gopalswamy2001predicting}; Marqu{\'e} et al. \citeyear{marque2006solar}; Gopalswamy et al. \citeyear{gopalswamy2008radio}; Swalwell et al. \citeyear{swalwell2017solar}; Gopalswamy et al. \citeyear{gopalswamy2017hierarchical}; Cliver \& D'Huys \citeyear{cliver2018size}; Rotti \& Martens \citeyear{rotti2023analysis}).

The Sun releases ions of about tens of Mega electron-Volt (MeV) and more in many SEP events. The motion of such highly charged particles is dictated by magnetic field lines (Reames \citeyear{reames2013two}). To be geo-effective, SEPs should reach the near-Earth space on a magnetically well-connected path (Cane et al. \citeyear{cane1988role}). Generally, it is understood that the eruptions at the western side of the Sun have a higher probability of SEPs being geo-effective due to the spiral structure of the interplanetary magnetic field lines, known as the Parker spiral (Parker \citeyear{parker1965dynamical}; Reames \citeyear{reames1999particle}). Such Earth-bound SEPs are hazardous on many levels. The impacts include severe technological (Smart \& Shea \citeyear{smart1992}) and biological effects on various economic scales (Schrijver \& Siscoe \citeyear{schrijver2010heliophysics}). Although Earth's magnetic field provides us a protective shield from the energetic particles and filters them out from reaching the ground, they can be life threatening for humans on missions outside of the Earth's magnetosphere and aircraft travels along polar routes (Beck et al. \citeyear{beck2005tepc}; Schwadron et al. \citeyear{schwadron2010earth}). For instance, long-lasting strong SEP events pose a radiation hazard to astronauts and electronic equipment in space (Jiggens et al. \citeyear{jiggens2019situ}). In addition, the particles getting diverted to Earth's magnetic poles disturb the ionosphere's upper layers, causing disruption to high-frequency radio, GPS, and satellite communications (Desai \& Giacalone \citeyear{desai2016large}).

Over the last four decades, great progress in space exploration has provided near-continuous observations of solar activity from a fleet of advanced space-based satellites (Usoskin \citeyear{usoskin2017history}). These observational data should be analyzed in operational contexts to mitigate SWx effects on our human explorers and technological systems (Jackman \& McPeters \citeyear{jackman1987solar}). Therefore, we require robust tools and systems to forecast eruptive event occurrences such as SEPs and send warning signals before the event. Researchers across the globe have been focusing on implementing a variety of model-driven techniques for predicting SEP events, mostly concentrating on predicting the peak flux-related characteristics. To forecast SEP event occurrences, physics-based and data-driven statistical models have been designed based on the parameters of parent solar eruptions such as SFs and CMEs (Van Hollebeke et al. \citeyear{1975van}; Posner \citeyear{posner2007up}; Kahler et al. \citeyear{2007kahler}; Balch \citeyear{balch}; Laurenza et al. \citeyear{laurenza2009technique}; {N{\'u}{\~n}ez} \citeyear{2011nunez}; Falconer et al. \citeyear{falconer2011tool}; Dierckxsens et al. \citeyear{dierckxsens2015relationship}; Anastasiadis et al. \citeyear{anastasiadis2017predicting}; Alberti et al. \citeyear{alberti2017solar}; Papaioannou et al. \citeyear{papaioannou2018nowcasting}).

Machine learning (ML) methods have been at the forefront of SEP event forecasting in the last decade due to their success in many other areas of research and operations (Boubrahimi et al. \citeyear{boubrahimi2017prediction}; Swalwell et al. \citeyear{swalwell2017solar}; Engell et al. \citeyear{engell2017sprints}; Camporeale \citeyear{enrico2019}; Aminalragia-Giamini et al. \citeyear{aminalragia2021solar}; Lavasa et al. \citeyear{lavasa2021assessing}; Whitman et al. \citeyear{WHITMAN2022}; Hosseinzadeh et al. \citeyear{hosseinzadeh2024improving}). Our previous work (Rotti, Aydin \& Martens \citeyear{rotti2024}) implemented time series classifiers on a multivariate time series (MVTS) constituting solar X-ray and proton fluxes data set of SEP events and showed that the interval-based supervised time series forest (STSF) model is more efficient in the classification of strong SEP events. Possible areas of improvement here are to consider an actual class imbalance scenario, that is, including the phases of non-SEPs and comparing with baseline time series classification algorithms for the task at hand. Expanding on these ideas in the present work, we implement STSF, summary statistic classifier (SSC) and one-nearest neighbor (1NN) classifiers on an extended data set constituting SEP-quiet periods and analyze their performances. With that, we compare the results and establish a proof of concept of the model to be implemented in a near-real-time (NRT) scenario. That is, we aim to build a high fidelity (low-risk) and interpretable short-term (low lead time) predictive model for reliable SEP event forecasting systems suitable to operational standards.

SEPs observed in near-Earth space are rare events compared to major flares and CMEs. In addition, the occurrences of SEPs are dependent on precursor solar eruptions (Cliver et al. \citeyear{cliver2022extreme}; Rotti \& Martens \citeyear{rotti2023analysis}). Thus, it makes physical sense to build SEP predictive models based on the parameters of parent solar events. However, in the present work, we consider only SFs and defer to the use of CME data due to their low data quality, such as the projection effects and difficulty of tracking them as 3d objects.

The rest of the paper is organized as follows: Section \ref{sec:data} provides information about our data set and data preparation steps used in this work. Section \ref{sec:methods} presents our research methodology, including the descriptions of the time series classification models and data partitioning scheme. Section \ref{sec:results} discusses the training and validation phases of the model and presents the experimental evaluation framework. Lastly, Section \ref{sec:conclusion} provides a summary of our work and future endeavors.

\section{Data} \label{sec:data}
In this work, we implement MVTS data integration, model training and selection pipeline toward building an efficient architecture for short-term SEP event forecasting. We consider a binary classification task between strong SEP and weak or non-SEP events. In the present framework, we consider two classes, ``SEP'' and ``SEP-quiet,'' where SEP-quiet periods include smaller proton flux increases, as well as the flares that did not lead to SEP events near Earth. With that, we maintain a natural class imbalance between the occurrence and non-occurrence of strong SEP events. Our methodology for the creation of SEP-quiet samples will be presented in the next Section. The strong and weak SEP samples in our work come from the Geostationary Solar Energetic Particle (GSEP) events data set (Rotti et al. \citeyear{rotti2022}), which is discussed later in Section \ref{sec:gsep}. 

\subsection{SEP-quiet periods}
The National Oceanic and Atmospheric Administration (NOAA) has been operating the Geostationary Operational Environmental Satellite (GOES) series in geosynchronous orbits since 1976 (Sauer \citeyear{sauer1989sel}; Bornmann et al. \citeyear{bornman}). The space environment monitor (SEM) onboard GOES missions measures solar radiation in the X-ray and EUV region and the in-situ magnetic field and energetic particle environment (Grubb \citeyear{grubb}). In this work, we utilize the GOES solar X-rays (Garcia \citeyear{garcia1994temperature}) and energetic proton fluxes (Onsager et al. \citeyear{onsager1996operational}). We create our negative samples using ``SEP-quiet'' periods in two stages: (i) obtaining lists of SF and SEP events and (ii) gathering the corresponding GOES flux measurements for each event. We note that we take into account the scaling factors initially implemented by NOAA in the GOES-08 through -15 X-ray measurements. Towards that, we updated our data set by removing the necessary scaling factors as per the NOAA guidelines\footnote{GOES XRS scaling factors: \url{https://ngdc.noaa.gov/stp/satellite/goes/doc/GOES_XRS_readme.pdf}} to get true X-ray fluxes.

NOAA-GOES has traced flaring activity since the beginning of 1974 and offers a catalog\footnote{The NOAA flare list: \url{https://www.ngdc.noaa.gov/stp/space-weather/solar-data/solar-features/solar-flares/x-rays/}} with spatial and temporal specifications, flare magnitude, and associated active region (AR) information. However, some flare locations and/or AR numbers are missing from the GOES catalog or probably have known errors (Milligan \& Ireland \citeyear{milligan}; Angryk et al. \citeyear{angryk2020mvts}; Rotti et al. \citeyear{rotti2020}). NOAA categorizes SFs based on soft X-ray peak flux in the wavelength range 1–8 {\AA}. Flare classes from least to most intense are labeled as A, B, C, M, and X, where, each category indicates an increase in flare intensity on a logarithmic scale.

\begin{figure}[ht!]
\plotone{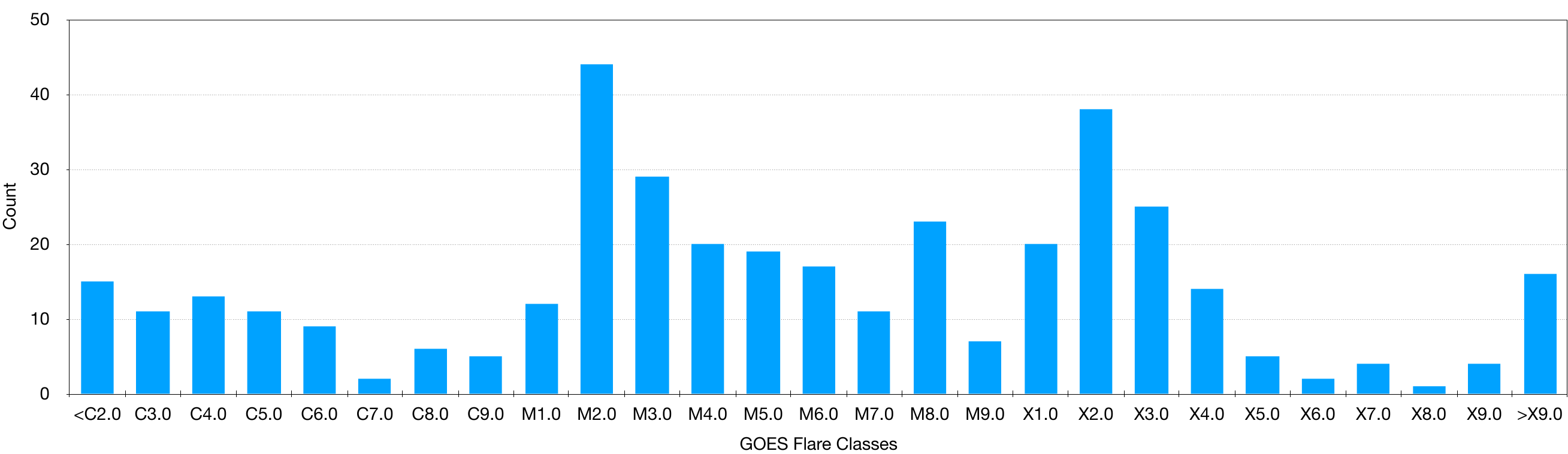}
\caption{Distribution of soft X-ray flare peak intensities based on GOES flare classification for SEP-associated flares in the GSEP data set. There are a total of 383 SEP with flare class information, of which 59 events have flare magnitudes less than GOES-class C6.0.
\label{fig:flare_hist}}
\end{figure}

\subsection{Non-SEP samples}
\begin{figure*}[ht!]
\gridline{\fig{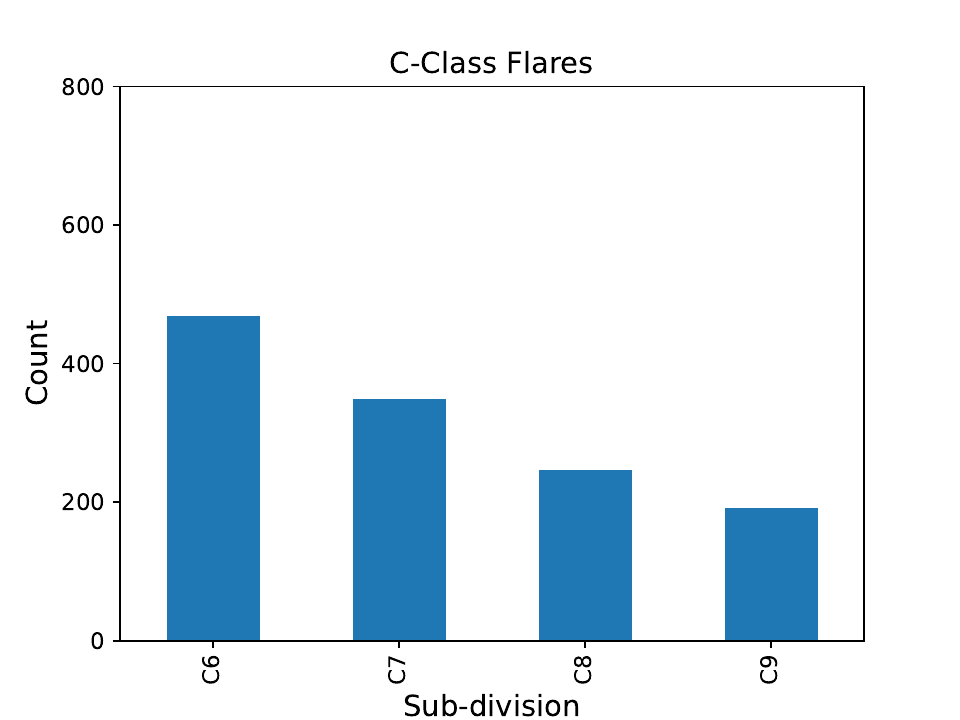}{0.48\textwidth}{(a)}
          \fig{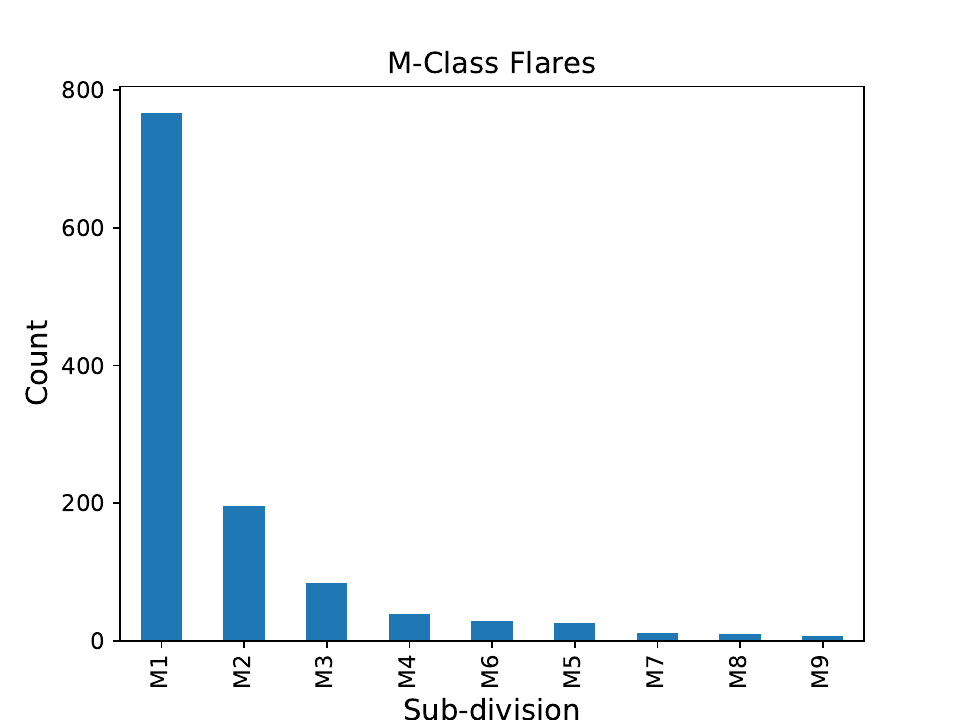}{0.48\textwidth}{(b)}
          }
\gridline{\fig{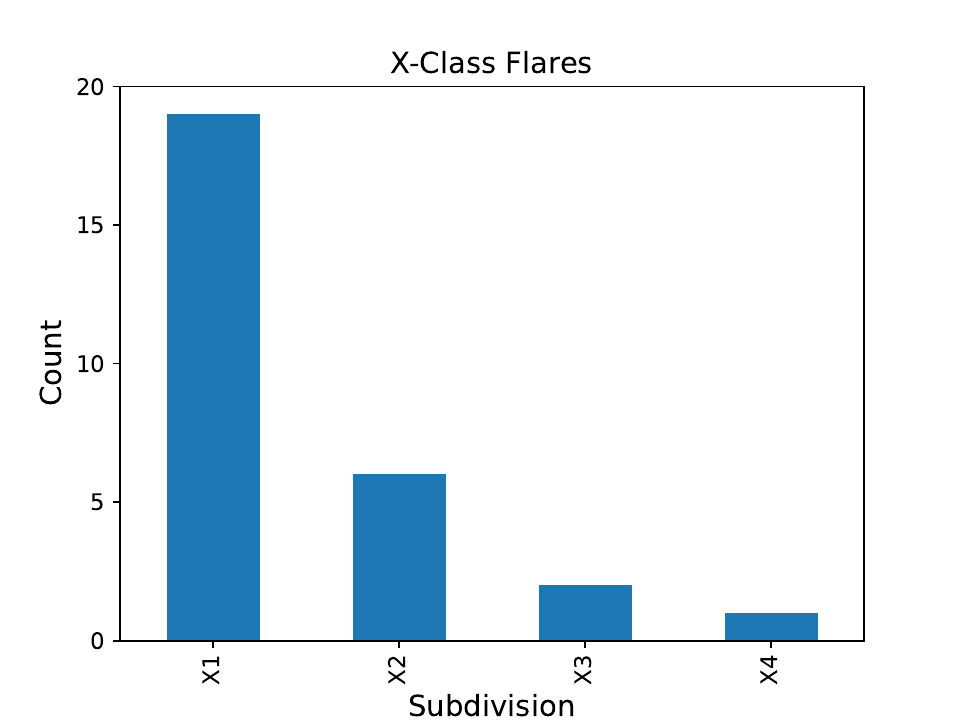}{0.48\textwidth}{(c)}
          }
\caption{Distribution of the GOES subdivisions of (a) C- (b) M- and (c) X-class flares present on our ``SEP-quiet'' periods.
\label{fig:class_dist}}
\end{figure*}

We select our non-SEP samples such that there are no enhancements or relatively no enhancements in the GOES proton fluxes following flaring episodes. In other words, we identify large flares that do not lead to significant variations in the GOES $\geq$ 10MeV proton fluxes relative to the background levels. This step helps us increase the sample size over the previous study. At first, we chose M and X-class flares from the NOAA flare list between 1986 and 2018 that have not been associated with strong SEP events. Then, to add a few more negative samples, we extend our lower limit of a ``non-SEP'' flaring intensity using a data-driven approach. The GSEP data set consists of carefully identified SF information for 383 associated SEP events. In that, we see a distribution of strengths amongst 59 SEP events associated with $<$C6.0 flares, where 38 are weak events and 21 are strong events. Nonetheless, we inspect all of the SEP-associated flares as shown in Figure \ref{fig:flare_hist} and  find that $<$C6.0 flares correspond to the bottom 15th percentile.  Hence, we consider C6.0 as our bottom threshold for flares. Then, we add all flares from M1.0 to C6.0 to our earlier flare list and obtain a total of 7,981 flares. We filter the preliminary negative-sample flares in two steps: (1) we do not consider flares during an ongoing SEP event (either rise or decay) and (2) we remove all consecutive flares within 11 hr following the onset of the flare at consideration. With this filtering step, our final count of non-SEP flares with an intensity $\geq$C6.0 reduced to 2,460. In Figure \ref{fig:class_dist}, we show the distribution of the number of ``SEP-quiet'' flares in each sub-divisions of GOES (\ref{fig:class_dist}a) C-, (\ref{fig:class_dist}b) M-, and (\ref{fig:class_dist}c) X-class.

Whitman et al. (\citeyear{WHITMAN2022}) recently organized an SEP model validation (SEPVAL) challenge in 2023\footnote{SEPVAL Challenge: \url{https://ccmc.gsfc.nasa.gov/community-workshops/ccmc-sepval-2023/}}. For this, two lists of 33 SEP and 30 non-SEP events from solar cycles (SCs) 24 and 25 have been provided. In this regard, we cross-verified our non-SEPs with the SEPVAL's list of 15 non-SEP events for cycle 24. We do not include any (SEP or non-SEP) events from SC25 in the present work to make our analysis comparable with earlier studies.

\subsection{Time Series Slicing}
To generate our dataset, we have interpolated and integrated particle and X-ray flux data from several GOES missions. First, we use the one-minute averaged X-ray (1–8 {\AA}) fluxes measured by the X-ray sensor (XRS) onboard GOES. The archived data is available online from NOAA's website\footnote{GOES-XRS: \url{https://www.ncei.noaa.gov/data/goes-space-environment-monitor/access/avg/}}. In addition to X-ray data, we use the following five-minute averaged integral proton fluxes measured by the SEM suite: (1) E$\geq$10 MeV fluxes corresponding to P3, (2) E$\geq$50 MeV fluxes corresponding to P5, and (3) E$\geq$100 MeV fluxes corresponding to P7. We interpolate the proton fluxes to one-minute averages and retain the time stamp of every first measurement. The rationale here is to capture the flare dynamics that occur over a few minutes. In the present work, we follow the same procedure as discussed in Rotti et al. (\citeyear{rotti2024}) to generate our event slices for the additional negative samples and ensures that our analysis and models' predictive capability are relevant and comparable across different SEP event prediction tasks.

We consider flares as the precursor to the onset of an SEP event. In our earlier statistical study on strong and weak SEP events (Rotti \& Martens \citeyear{rotti2023analysis}), we find that 77($\pm$4)\% of SEP event onsets occur within 11 hrs after the associated flaring eruption. However, sampling flares from the onset will likely lead the classifier to memorize the pattern of finding a spike at the start of the MVTS instance. A simple solution to randomize the sampling is to include 1 hour of data prior to the flare onset because two to three flares may occur within an hour of observation in many instances. With that consideration, the classifiers become free from memorizing to find a spike at the start of the data or at the end of 60 data points corresponding to one hr. We accommodate both these aspects while creating our non-SEP time series sampling because our goal is to develop a highly accurate model predicting strong SEP events. As reported in our previous work (Rotti et al., \citeyear{rotti2024}), we note that our modeling schema can predict well-connected events in addition to strong proton enhancements. A caveat/drawback of the model would be to classify events that fluctuate around the 10 pfu threshold.

\begin{figure}[ht!]
\plotone{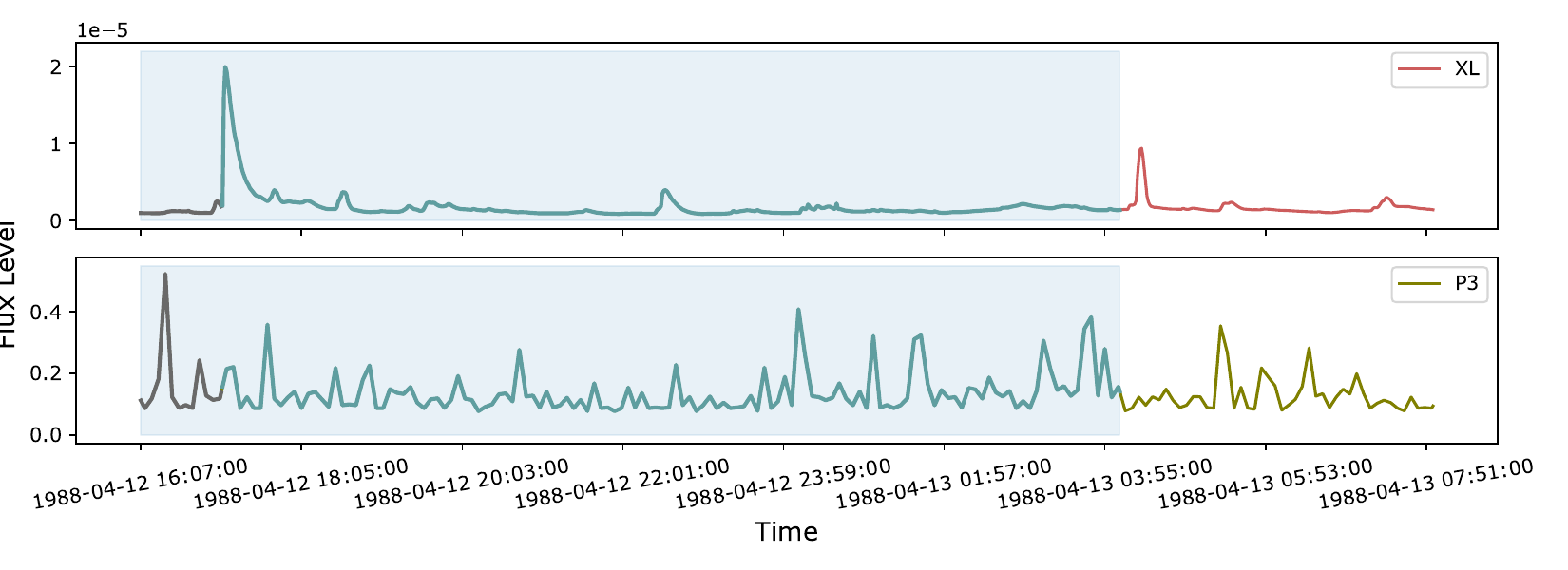}
\caption{An example of the sampling technique for the non-SEP samples in our data set. We show the XL and P3 fluxes for representation only. The blue-shaded region in the plot shows the 12-hour length of the input time series of our sample. Here, the grey-shaded line on the left corresponds to the one hour of fluxes prior to the ($\geq$C6.0) flare onset, while the rest of the time series within the blue region covers 11 hours after the onset.
\label{fig:neg_schema}}
\end{figure}

Based on our filtered list of 2,460 flares for negative samples, we consider 1 hour of data (all four fluxes mentioned above) before the flare onset and 11 hr afterward for each flaring event (as shown in Figure \ref{fig:neg_schema}). Therefore, the total length of each MVTS corresponds to 12 hr. As our final data set comes with a cadence of one minute, each MVTS is a size-720 sequence of soft X-ray and integral proton fluxes. All time series data was cleaned for data gaps with linear interpolation and spike removal by replacing what would be considered an outlier (spike) with an interpolated value from moving averages. We note here that the fluxes are predominantly derived from primary GOES missions. Nonetheless, when there are ambiguous measurements in the primary satellite, we consider data from the secondary satellite of the corresponding period of observation. 

Because we are focusing on short-term predictions, we explore multiple prediction windows: 5-, 15-, 30-, 45-, and 60-min. In each case, we exclude the respective minutes of observations from the 720-size MVTS at the end by maintaining a constant 11 hr observation window as input to the model. Such a continuous and consistent observation window is needed for training a predictive model to be established in a near-real-time operational environment. The prediction window indicates the corresponding time in the future at the end of the observation window for when the model output is valid.

\subsection{GSEP data set}\label{sec:gsep}
Our primary source of SEP events comes from the recently published open-source GSEP data set (Rotti et al. \cite{gsep_2022}). It consists of a carefully vetted SEP events catalog with associated parent eruption metadata and time series slices. The data set comprises 433 (- 244 strong and - 189 weak) SEP events covering solar cycles 22 to 24. According to the NOAA Space Weather Prediction Center (SWPC), a strong SEP event is defined as having proton intensities $I_{P}$ $\ge$10 pfu (1 pfu = 1 particle per cm$^2$.s.sr) in the GOES five minutes averaged E$\ge$10 MeV integral energy channel for at least three consecutive readings (Bain et al. \citeyear{bain2021summary}). In the GSEP data set, ``weak'' SEP events correspond to an enhancement in proton intensities in E$\ge$ 10 MeV channel whose peak flux measurements do not reach to 10 pfu. The description of the data set and its development can be found in Rotti et al. (\citeyear{rotti2022}) and Rotti \& Martens (\citeyear{rotti2023analysis}).

\section{Methodology} \label{sec:methods}
\begin{figure}[ht!]
\plotone{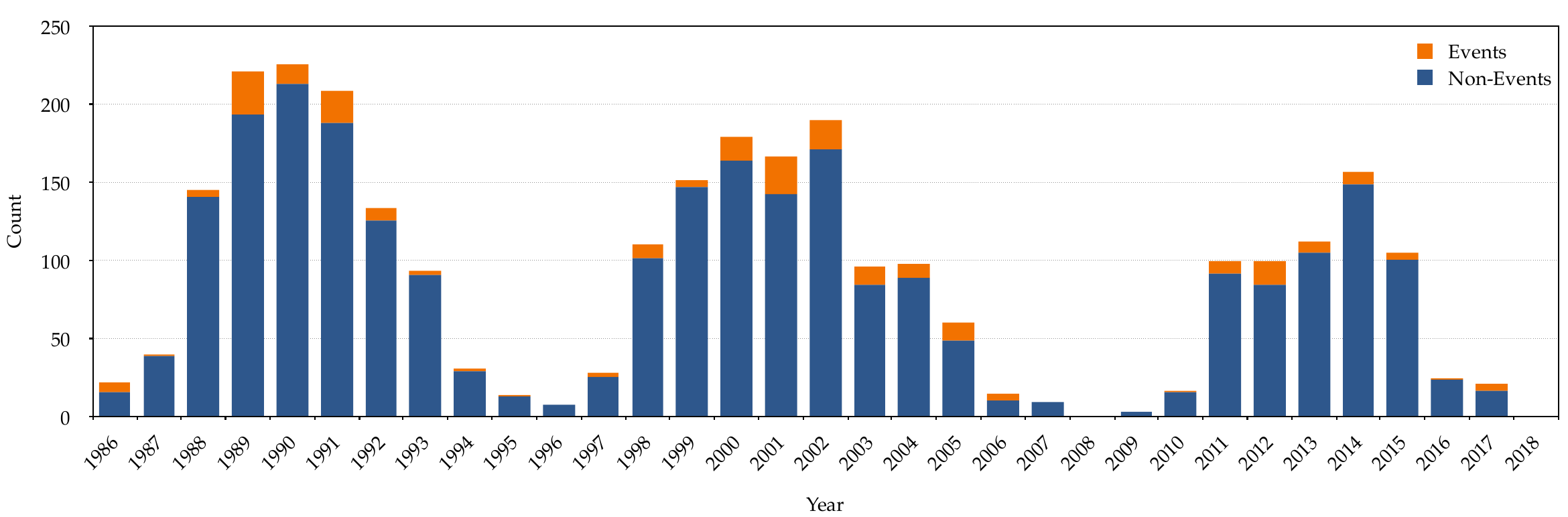}
\caption{The distribution of SEP and non-SEP samples in our data set between 1986 and 2018. We split our data set into three parts of non-overlapping years: Partition 1 = 1986 - 1992; Partition 2 = 1993 to 2002; and Partition 3 = 2003 to 2018. Here, each split consists of 998, 974 and 921 samples, respectively. The data set has a cumulative class-imbalance ratio of $\approx$ 1:11. In the legend, ``Events" are defined as those SEPs crossing the SWPC threshold of 10 pfu in the E$\geq$10 MeV channel, and ``Non-Events" consists of both weaker enhancements $<$10 pfu or no enhancements. 
\label{fig:bar}}
\end{figure}

We consider the term `SEP events' analogous to solar protons events (SPEs). While variations exist, positive labels are usually associated with the occurrence of strong SEPs based on the integral proton fluxes recorded by the GOES-P3 channel crossing the 10 pfu threshold. As mentioned earlier, we have 189 weak SEP events in the GSEP data set and we add 2,460 non-SEPs to increase the negative samples to 2,649. For the positive class, we have 244 strong SEP events. The total size of the data set is 2,893 samples with a class-imbalance ratio of $\approx$ 1:11. Figure \ref{fig:bar} shows the total number of samples in our data set from 1986 to 2018. Here, `Events" are defined as those SEPs crossing the SWPC threshold of 10 pfu in the E$\geq$10 MeV channel, and ``Non-Events" constitute both weaker enhancements of $<$10 pfu and no enhancements. In Section \ref{sec:tsc}, we briefly explain the method of time series classification implemented in the present work.

\subsection{Time Series Classification}\label{sec:tsc}
We examine the performance of time series classifiers in the framework of a binary classification task for predicting SEP events. Time series classification uses supervised ML to analyze labeled classes of time series data and then predicts the class to which a new instance belongs. This is important in SWx predictions, where observational data is analyzed to support NRT operations and classification accuracy becomes critical. Furthermore, for short-term predictions, the algorithm must be highly accurate and robust to be useful. In our previous work (Rotti et al. \citeyear{rotti2024}), we showed that an ensemble framework of interval-based classifiers achieved high performance in classifying between strong and weak SEP events. 

In the present work, we implement three time series classifiers and compare their performance to establish certainty in our SEP event prediction system. The motive is to see whether the input time series span leads to a strong SEP event (positive class) or not (negative class). We implement the same column ensemble schema presented in our previous work (Rotti et al. \citeyear{rotti2024}), where the ensemble estimator allows multiple feature columns of the input to be transformed separately. The statistical features generated by each classifier on samples of the original time series are ensembled to create a single output. Each feature is assigned a score that indicates how informative it is towards predicting the target variable (Hansen \& Salamon \citeyear{hansen1990neural}; Schapire \citeyear{schapire1990strength}; Arbib \citeyear{arbib2003handbook}). The predictions from individual classifiers are then aggregated with equal votes using prediction probabilities. We will give a brief overview of our models in the following sections.

\subsection{Summary Statistic Classifier}\label{sec:ssc}
The feature-based summary statistic classifier is the first of the two baseline models we have used in this work. This classifier applies a time series feature transformation to an entire univariate time series using simple summary statistics and builds a random forest using these summary statistics. The summary statistics extracted are mean, standard deviation, minimum, maximum, and quantile values (at 0.25, 0.5, and 0.75, corresponding to first and third quartiles and median). The tabulated summary statistics are then fed to a random forest classifier for training a univariate time series classifier. For applying this method to our multivariate time series classification task, we employed a late fusion approach, where univariate time series classification models trained on individual parameters are ensembled based on a majority voting schema.

\subsection{One Nearest Neighbor}\label{sec:1nn}
The similarity-based (nearest neighbor) time series classification is one of the most popular classification methods due to its simplicity, nonparametric nature, potential for interpretability, and flexibility in utilizing different similarity schemes/metrics. It assigns a label for a new instance based on the target value of k-most similar instances in the training dataset. The similarity is assessed based on the inverse of a designated time series distance measure and utilization of elastic distance measures allows capturing the local and global similarity patterns (Faouzi \citeyear{faouzi2022time}). In this work, we implement a k-nearest neighbor (kNN, with k=1) classifier with Dynamic Time Warping (DTW) distance, which is a commonly used elastic measure (Sakoe \& Chiba \citeyear{1163055}). This 1NN is a baseline classifier for time series classification and similar to our earlier approaches, we use 1NN with DTW distance to issue predictions. 

\subsection{Supervised Time Series Forest Classifier}\label{sec:stsf}
The supervised time series forest (STSF; Cabello et al. \citeyear{cabello2020fast}) classifier comes under the category of interval-based models, which uses three representations (time, frequency, and derivative) of the input time series to extract features from a supervised selection of intervals. That is, for each representation, STSF builds an ensemble of decision trees on intervals selected through a supervised process wherein the algorithm finds the discriminatory intervals. STSF computes the region of interest (ROI) to highlight the location of discriminatory intervals, which are the intersected regions of such intervals. It extracts seven features, namely mean ($\mu$), standard deviation ($\sigma$), slope (\textit{m}), median, interquartile range (IQR), minimum value and maximum value, from each interval. The ranking of the interval feature is obtained by a scoring function that indicates how well the feature separates one class of time series from the other classes. The final set of intervals is obtained in a top-down approach to represent the entire series. The feature set is concatenated to form a new data set upon which decision trees are built. The final output is based on majority voting of averaged probability estimates of the individual estimators in the ensemble.

\subsection{Data Partitions}\label{sec:partitions}
As we are implementing a classification task with target labels, we split our data set into three sets: the training set, the holdout or validation set, and the test set. This partitioning schema is relevant to estimate the predictive capacity of the model on different aspects. For instance, the validation set is used in the present work to obtain the optimal threshold of classification for each model, estimate the skills scores and compare their performance. Furthermore, we use the unseen test set to sanity-check the best model's predictive capabilities.

We split our data set into three sets on non-overlapping years (1986-1992, 1993-2002 and 2003-2018) such that there are 998 samples for training, 974 samples for validation and 921 samples for testing the model. Here, each partition consists of a similar number of positive samples ($\approx$ 80) and has similar class imbalance ratios, that is, $\approx$ 1:11. Table \ref{tab:dp} shows the number of data samples in each partition in relation to the binary class labels.

\begin{deluxetable*}{r|c|c|c|}[ht!]
\tablenum{1}
\tablecaption{Partitioning strategy of our data set. \label{tab:dp}}
\tablewidth{0pt}
\tablehead{
\textbf{}   &  {\textbf{Training}} & \textbf{Validation} &{\textbf{Test}}
}
\startdata
Positive & 80 & 80 & 84\\
|||| &|||&|||&|||\\
Negative & 918 & 894 & 837\\
\enddata
\tablecomments{Number of samples in each partition corresponding to the binary target labels. Here, labels correspond to a yes (positive) or no (negative).}
\end{deluxetable*}

\vspace{-3em}
\section{Results} \label{sec:results}
In this work, we aim to demonstrate the robustness of the methodology for SEP event forecasting that was proposed in our previous work (Rotti et al. \citeyear{rotti2024}) and provide a proof of concept on its efficiency in generating short-term predictions. For this purpose, we compare three classifiers (SSC, 1NN and STSF) and analyze their performances. We construct a binary classification task such that the `positive' class consists of strong SEP events while the `negative' class contains the rest. We use the training set to train the model and perform a grid search for hyperparameter optimization of our classifiers. The best hyperparameters for STSF and SSC were on the default model settings with the number of estimators = 200.

We explore the model capabilities for different short-term prediction windows: 5-, 15-, 30-, 45-, and 60-minutes. We use the following terms henceforth to identify different prediction windows that act as lead times for model predictions, respectively: T$_{5}$, T$_{15}$, T$_{30}$, T$_{45}$ and T$_{60}$. In each experiment, we consider the respective prediction window consisting of a constant 11-hour observation window from our MVTS samples before training the model. We implement multiple metrics (see Section \ref{sec:metrics}) to estimate the predictive/forecast skill of our models.

\subsection{Metrics} \label{sec:metrics}
We consider a 2 $\times$ 2 contingency table with the following elements: true positive (TP), true negative (TN), false positive (FP), and false negative (FN). Here, TP indicates the number of correctly predicted strong SEP events (positive class), while TN represents the number of correctly predicted weak and non-SEP events (negative class). FP corresponds to the number of negative classes predicted as positives (false alarms), while FN corresponds to the number of positive class labels predicted as negatives (misses).

On a simple scale, statistical metrics such as precision (Eq. \ref{sc:pre}) and recall (Eq. \ref{sc:rec}) have been traditionally used to assess classifier performances. Precision for the positive class is used to evaluate the model's correct prediction with respect to the false alarms. Recall for the positive class characterizes the ability of the classifier to find all of the positive cases.

\begin{equation} \label{sc:pre}
Precision = \frac{(TP)}{(TP + FP)}
\end{equation}

\begin{equation} \label{sc:rec}
Recall = \frac{(TP)}{(TP + FN)}
\end{equation}

Focusing on the importance of positive classes, we consider the F$_{1}$-score for the positive class that can be estimated as the harmonic mean of precision and recall as shown in Equation \ref{sc:f1}. It ranges between 0 and 1 such that scores closer to 1 indicate the model to be better. 

\begin{equation} \label{sc:f1}
F{_1} = 2 \times \frac{(Precision \times Recall)}{(Precision + Recall)}
\end{equation}

To understand the overall model performance, we use a `weighted' average for the F$_{1}$-score (shown in Equations \ref{sc:f1_wt} and \ref{sc:wt}), which computes the score for each target class and uses sample weights that depend on the number of instances in that class while averaging. Here, i is the number of target classes in the data set, which is two in the present work.

\begin{equation} \label{sc:f1_wt}
F_{1_{weighted}} = \sum_{i=1}^{N} w{_i} \times F_{1_i}
\end{equation}

\begin{equation} \label{sc:wt}
w{_i} = \frac{\text{Number of samples in class \textit{i}}}{\text{Total number of samples}}
\end{equation}

We implement true skill statistics (TSS; Woodcock \citeyear{woodcock_evaluation_1976}; Dann \citeyear{daan1985forecast}) to account for the false positive rate comparing the difference between the probability of detection and the probability of false detection as shown in Equation \ref{sc:tss}. TSS ranges from -1 to +1, where the latter indicates a perfect score. TSS $\leq$0 indicates it is worse than a random classification.

\begin{equation} \label{sc:tss}
TSS = \frac{(TP \times TN) - (FP \times FN)}{(TP + FN) \times (FP + TN)}
\end{equation} 

Furthermore, we consider the Heidke skill score (HSS; Heidke \citeyear{heidke_berechnung_1926}) that measures the improvement of the forecast over a random prediction as defined in Equation \ref{sc:hss}. HSS with 1 indicates perfect performance and 0 indicates no skill. A no-skill means the forecast is not better than a random binary forecast based on class distributions.
\begin{equation} \label{sc:hss}
HSS = \frac{2 \times ((TP \times TN) - (FP \times FN))}{((TP + FN) \times (TN + FN)) + ((FP + TN) \times (FP + TP))}
\end{equation}

The Gilbert Skill Score (GSS; Schaefer \citeyear{schaefer1990critical}) considers the number of hits due to chance, which is the frequency of an event multiplied by the total number of forecast events. This score formula is given by Equation \ref{sc:gss}. GSS ranges from -1/3 to 1. Here, 0 indicates no skill, while 1 is a perfect forecast.

\begin{equation} \label{sc:gss}
GSS = \frac{TP - (\frac{(TP + FN) \times (TP + FP)}{TP + FP + TN + FN})}{(TP + FP + FN) - (\frac{(TP + FN) \times (TP + FP)}{TP + FP + TN + FN})}
\end{equation}

Accounting for TN to assess the performance of a binary class problem is essential in our context. Hence, we also implement Matthew's correlation coefficient (MCC) as defined in Equation \ref{sc:mcc}. MCC ranges from -1 to 1 where 0 indicates no skill and 1 shows perfect agreement between predicted and actual values. MCC is a robust metric that follows a comprehensive and balanced strategy for using the contingency table elements. That is, a high MCC score is obtained only when the predictions proportionately account for the size of both positive and negative samples in the data set. This is important because popular scores such as F$_{1}$ fall short to account for class imbalance.

\begin{equation} \label{sc:mcc}
MCC = \frac{(TP \times TN) - (FP \times FN)}{\sqrt{(TP + FP) \times (TP + FN) \times (TP + FP) \times (TN + FN)}}
\end{equation}

\subsection{Validation set}
As mentioned in Section \ref{sec:partitions}, we use the second partition from our data set as a hold-out portion to estimate the optimal threshold for our classifiers. The classification threshold is the decision threshold that allows us to map the probabilistic output of a binary classifier to a binary category. In other words, it is a cut-off point used to assign a specific predicted class label for each sample. By default, the classification threshold in our models is 0.5. What that means is any prediction above 0.5 belongs to the positive class and that below 0.5 belongs to the negative class. However, 0.5 is not always optimal, especially when we have an imbalanced data set. Therefore, we identify a reliable threshold for the classifier that better discriminates between the two class labels. The first step is to derive the prediction probabilities from the validation set and extract the best trade-off between true positive rate (TPR) and false positive rate (FPR). In general, any binary classification model predicts mean probabilities for each input sample belonging to the positive class, where the prediction score from the classifier is greater than a parametrized threshold. Then, a classification threshold (from 0.0 to 1.0) is used to assign a binary label to the predicted probabilities. 

\begin{equation} \label{sc:J}
J = Sensitivity + Specificity - 1
\end{equation}

\begin{figure}[ht!]
\plotone{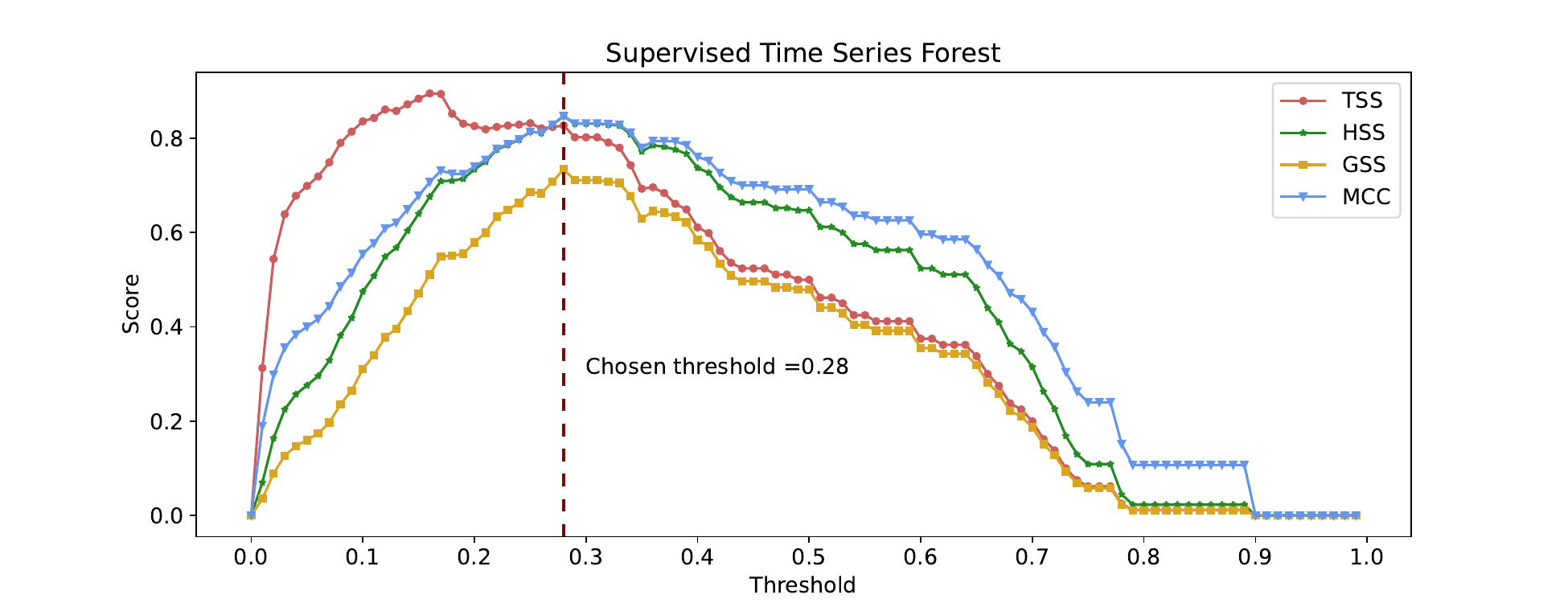}
\caption{The variation in skills such as TSS, HSS, GSS and MCC with respect to increasing the classification threshold for the STSF model on the validation/hold-out set at a prediction window of thirty minutes (T$_{30}$). The optimal threshold value for the model is inset in the plot.
\label{fig:comp}}
\end{figure}

\begin{figure*}[ht!]
\gridline{\fig{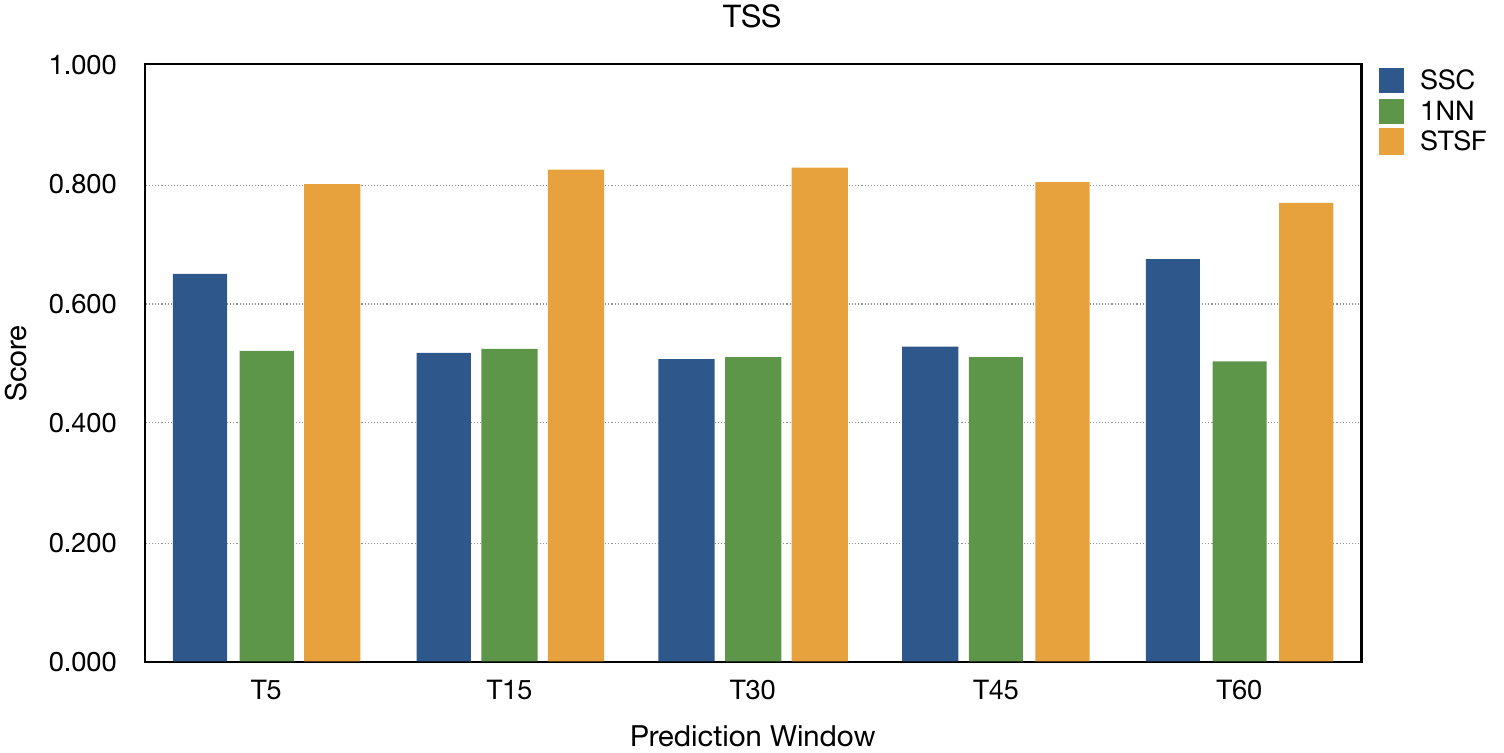}{0.45\textwidth}{(a)}
          \fig{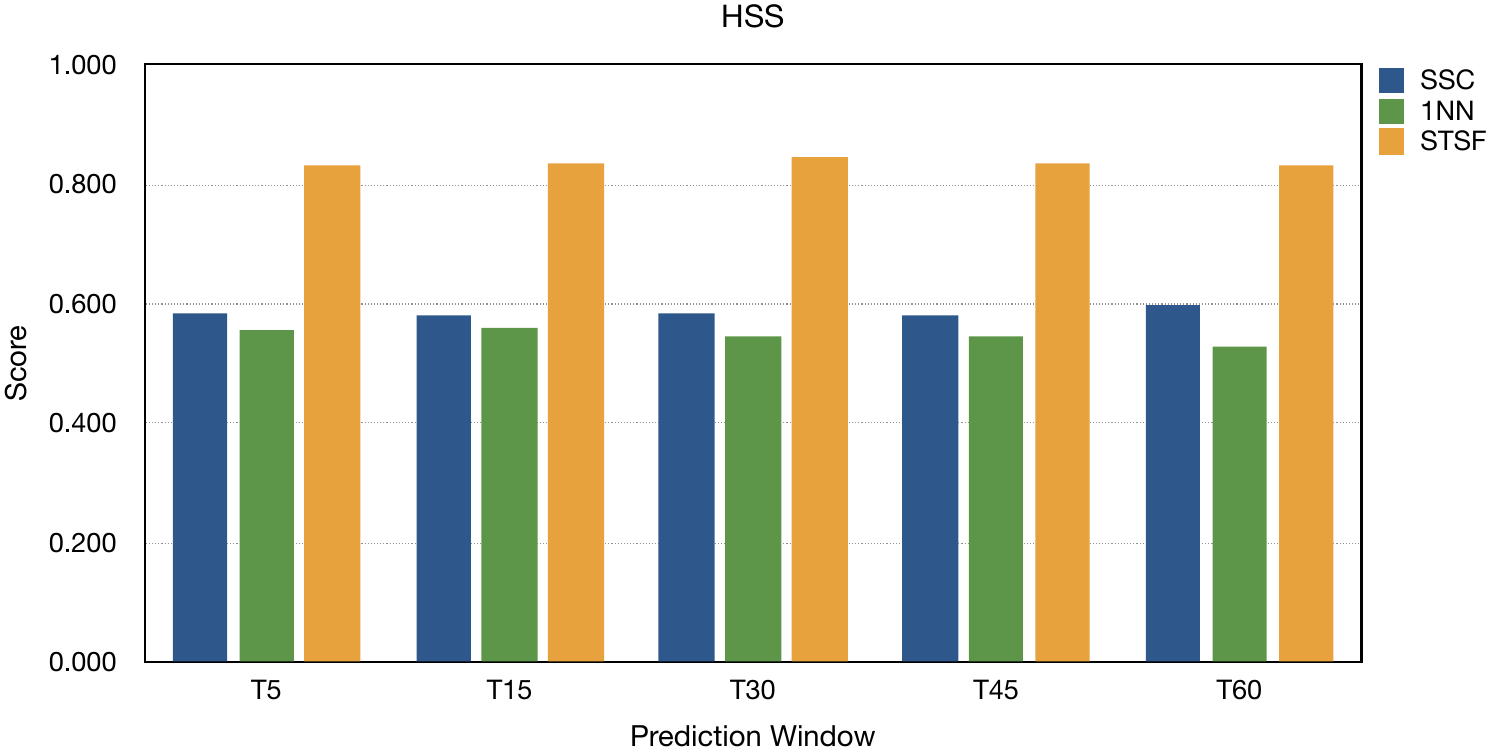}{0.45\textwidth}{(b)}
          }
\gridline{\fig{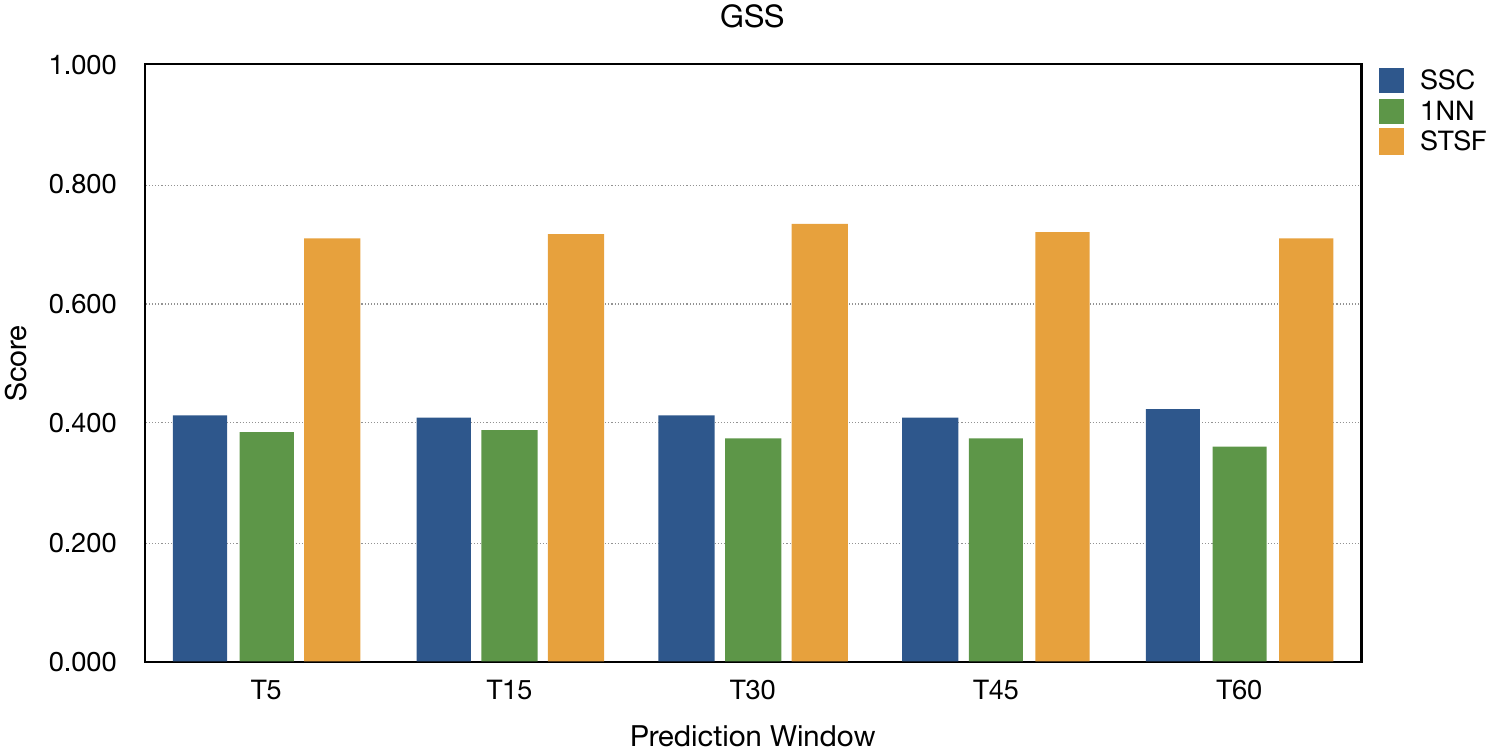}{0.45\textwidth}{(c)}
          \fig{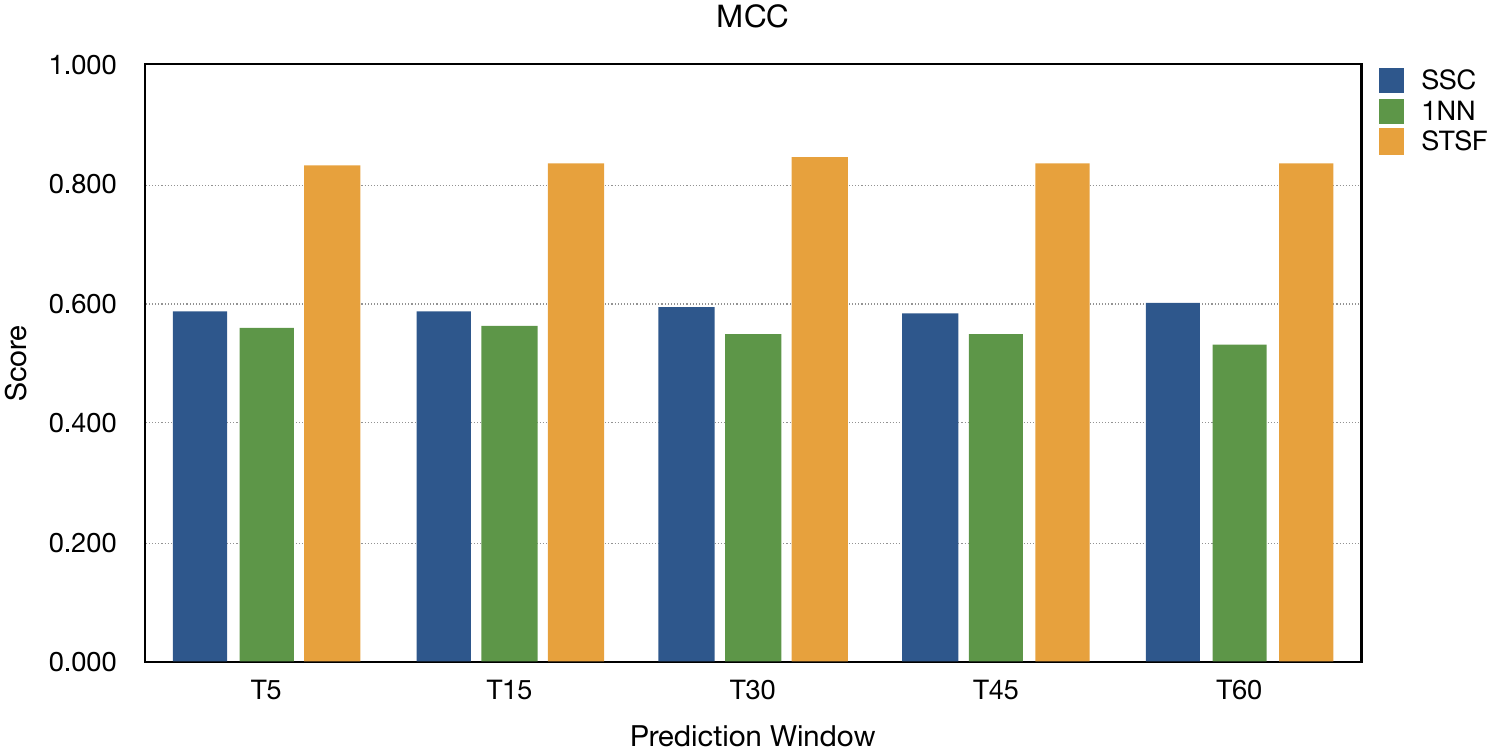}{0.45\textwidth}{(d)}
          }
\caption{Comparison of SSC, STSF and 1NN classifiers with respect to variation in skills such as (a) TSS, (b) HSS, (c) GSS and (d) MCC for prediction windows T$_5$, T$_{15}$, T$_{30}$, T$_{45}$, and T$_{60}$ on the validation set.
\label{fig:ss_comp}}
\end{figure*}

To find the optimal threshold that minimizes the difference between TPR and FPR, the Youden Index (J; Youden \citeyear{youden1950index}) is used in the literature as defined in Equation \ref{sc:J}. Here, sensitivity is the recall for the positive class and specificity is the recall for the negative class. J is the specialized version of TSS for the binary classification task. As a characteristic example, we show in Figure \ref{fig:comp} the effect of ``thresholding'' on the performance of STSF by visualizing the variations in the skills due to changing thresholds. Here, the estimated optimal threshold of STSF for T$_{30}$ on our data generation and sampling method is 0.28. We obtain optimal classification thresholds at all prediction windows for all the models and estimate the model skills. A comparison of the model performances is shown in Figure \ref{fig:ss_comp}.

\vspace{-1.5em}

Computationally, the run time of SSC was the fastest, followed by STSF and then 1NN. In terms of performance, 1NN falls behind both models but maintains relatively close values for all the skill scores in each prediction window with $\sigma$ $\approx$ 0.04. On the other hand, SSC shows a similar trend in its skills with $\sigma$ $\approx$ 0.01 except for TSS ($\sigma$ $\approx$ 0.07). Although SSC was a little faster, the performance marginally lags on our data set compared to STSF. For example, the average TSS for SSC is $\approx14\%$ lower than that for STSF. The STSF model performs exceptionally well compared to the other two baseline classifiers at all parameters. Specifically, the high scores at T$_{60}$ (e.g., TSS = 0.850) provide satisfactory reliability to our methodology. Table \ref{tab:valstsfcm} presents the contingency tables for all prediction windows corresponding to the optimal threshold of STSF on the validation set. The contingency tables for SSC and 1NN are provided in Tables \ref{tab:valssccm} and \ref{tab:val1nncm}, respectively, in the Appendix.

\movetabledown=3mm
\begin{longrotatetable}
\begin{deluxetable*}{ll|c|c|c|c|c|c|c|c|c|c|c|c|c|}
\tablenum{2}
\tablecaption{Contingency tables for the STSF model on the validation set. \label{tab:valstsfcm}}
\tablewidth{0pt}
\tablehead{
\textbf{} & \textbf{} & \multicolumn{2}{c|}{\textbf{T$_5$}} & \multicolumn{2}{c|}{\textbf{T$_{15}$}} & \multicolumn{2}{c|}{\textbf{T$_{30}$}} & \multicolumn{2}{c|}{\textbf{T$_{45}$}} & \multicolumn{2}{c|}{\textbf{T$_{60}$}} \\
\textbf{} & \textbf{} & \multicolumn{2}{c|}{Predicted} & \multicolumn{2}{c|}{Predicted} & \multicolumn{2}{c|}{Predicted} & \multicolumn{2}{c|}{Predicted} & \multicolumn{2}{c|}{Predicted} \\
 &  & Strong & Weak & Strong & Weak & Strong & Weak & Strong & Weak & Strong & Weak}
\startdata
 & Strong & 65 & 15 & 67 & 13 & 67 & 13 & 65 & 15 & 62 & 18 \\
True & ||| & ||| & ||| & ||| & ||| & ||| & ||| & ||| & ||| & ||| & ||| \\
 & Weak & 9 & 885 & 11 & 883 & 9 & 885 & 8 & 886 & 5 & 889 \\
\enddata
\tablecomments{Truth tables for the STSF model for different prediction windows. The first column is a shared entry of true labels against predictive labels for each experiment. The elements indicate the number of predictions with respect to the actual occurrences in the validation set. 
\newline
prediction windows are shown by a subscript to T. For example, T$_5$ indicates a prediction window of 5 min.
\newline
Model name: STSF - Supervised Time Series Forest}
\end{deluxetable*}
\end{longrotatetable}

\movetabledown=3mm
\begin{longrotatetable}
\begin{deluxetable*}{ll|c|c|c|c|c|c|c|c|c|c|c|c|c|}
\tablenum{3}
\tablecaption{Contingency tables for the STSF model on the test set. \label{tab:stsfcm}}
\tablewidth{0pt}
\tablehead{
\textbf{} & \textbf{} & \multicolumn{2}{c|}{\textbf{T$_5$}} & \multicolumn{2}{c|}{\textbf{T$_{15}$}} & \multicolumn{2}{c|}{\textbf{T$_{30}$}} & \multicolumn{2}{c|}{\textbf{T$_{45}$}} & \multicolumn{2}{c|}{\textbf{T$_{60}$}} \\
\textbf{} & \textbf{} & \multicolumn{2}{c|}{Predicted} & \multicolumn{2}{c|}{Predicted} & \multicolumn{2}{c|}{Predicted} & \multicolumn{2}{c|}{Predicted} & \multicolumn{2}{c|}{Predicted} \\
 &  & Strong & Weak & Strong & Weak & Strong & Weak & Strong & Weak & Strong & Weak}
\startdata
 & Strong & 71 & 13 & 73 & 11 & 75 & 9 & 73 & 11 & 72 & 12 \\
True & ||| & ||| & ||| & ||| & ||| & ||| & ||| & ||| & ||| & ||| & ||| \\
 & Weak & 6 & 831 & 6 & 831 & 8 & 829 & 6 & 831 & 6 & 831 \\
\enddata
\tablecomments{Truth tables for the STSF model for different prediction windows. The first column is a shared entry of true labels against predictive labels for each experiment. The elements indicate the number of predictions with respect to the actual occurrences in the test set. 
\newline
Prediction windows are shown by a subscript to T. For example, T$_5$ indicates a prediction window of 5 min.
\newline
Model name: STSF - Supervised Time Series Forest}
\end{deluxetable*}
\end{longrotatetable}

\subsection{Test set}\label{sec:testset}
In our previous work (Rotti et al. \citeyear{rotti2024}), we found that an ensemble of interval-based STSF classifiers is a highly efficient and optimal model to predict strong SEP events. STSF's approach to computing interval features in a supervised manner from three different time series representations by using robust statistics was a viable option for our problem specification. In this paper, we extended our modeling strategy by including non-SEP periods in our data set. Furthermore, we compared the performance of STSF and two baseline classifiers, namely SSC and 1NN. Based on the forecasting skill scores, we used the validation set from our data partitions to estimate the model performances. This resulted in STSF as the best model under all comparative paramters. We use the test set to further assess the model's capabilities. 

In Table \ref{tab:stsfcm}, we show the contingency or the confusion matrix elements derived for the STSF classifier on the test set for all prediction windows (T$_5$ to T$_{60}$) considered in this work. The increasing prediction window shows a relatively consistent type II error (false negative rate) except for T$_{30}$. That is, the number of false positives is lower compared to false negatives. A possible reason for the situation here is that several weak SEP events have proton fluxes closely fluctuating in the vicinity of the SWPC `S1' threshold. Detecting such patterns at high precision becomes relevant to reducing FNs.

\begin{figure}[ht!]
\plotone{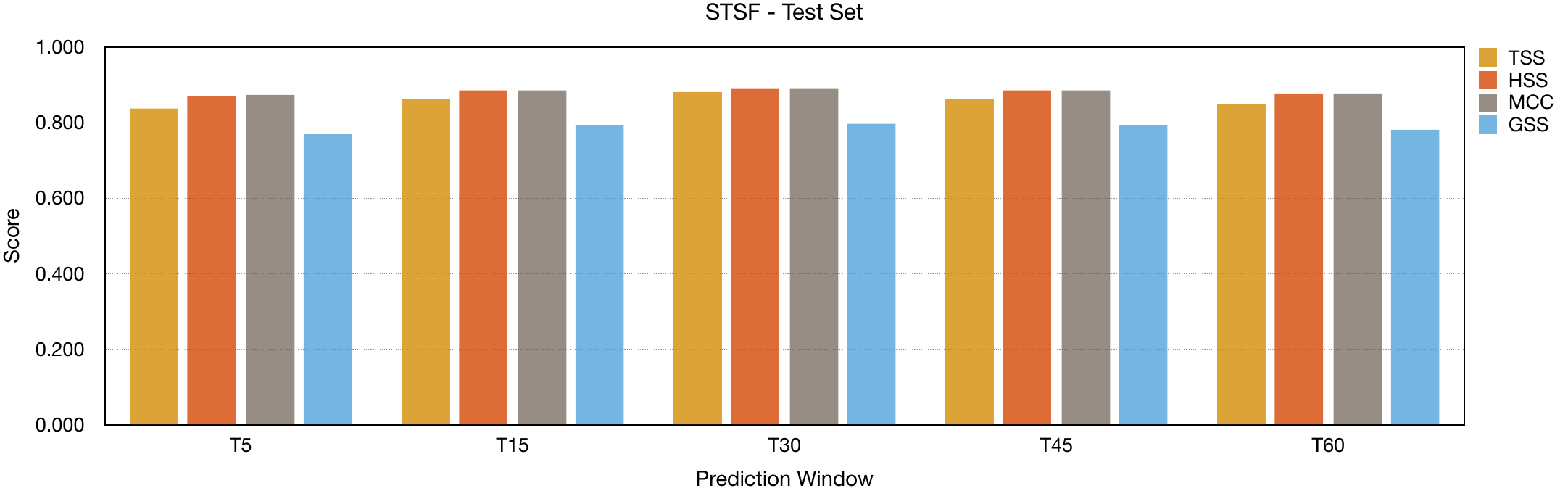}
\caption{Comparison of skill scores of STSF model on the test set at prediction windows T$_5$, T$_{15}$, T$_{30}$, T$_{45}$, and T$_{60}$.
\label{fig:sc_comp}}
\end{figure}

In Figure \ref{fig:sc_comp}, we compare the skill scores of STSF with respect to the different prediction windows considered in this work. The average TSS at all prediction windows is $\approx$ 0.86($\pm$0.02). Similarly, the HSS, MCC and GSS scores obtained are also high. There is a slight increment in the scores from T$_{5}$ to T$_{30}$, but reduces at T$_{45}$ and T$_{60}$. This slight variation is due to subtle changes in the optimal classification threshold $\approx$ 0.3($\pm$0.03) for STSF. The scores for T$_{60}$ only have a marginal reduction ($<$2\%) compared to T$_5$. Furthermore, there is a decrease of $\approx$ 7\%($\pm$2\%) in skill scores compared to Rotti et al. (\citeyear{rotti2024}). This is expected due to extended lead times considered with a larger class imbalance in the data set. Overall, the classification and computational efficiencies we obtain for STSF are highly satisfactory and viable. This means that our work adequately provides a proof of concept of our modeling architecture for short-term predictive capabilities up to a 60-minute window and can be further transformed into implementation for NRT operations.

\section{Conclusions} \label{sec:conclusion}
In this work, we approach the SEP event short-term prediction problem from a time series classification perspective. We primarily use the GSEP data set but extend our negative sample size by including ``SEP-quiet'' periods. That is, we consider $\geq$C6.0 flares that do not lead to SEP events observed near Earth. Our data set constitutes in-situ time series measurements from the NOAA-GOES missions for solar cycles 22 to 24. We utilize the long band (1–8Å) X-ray measurements from the XRS instrument and proton fluxes (P3, P5, P7) from the SEM instrument onboard GOES missions. Our data set has a total of 2893 samples, of which 244 are strong SEP events. We defined the target labels based on the SWPC `S1' threshold for an SEP event. Positive labels are strong events crossing the 10 pfu threshold in the 10 MeV proton channel, and negative otherwise.

Our goal is to develop robust models that can successfully find discriminants between SEP and SEP-quiet patterns. In our recent study (Rotti et al. \citeyear{rotti2024}), we found our methodology of using an ensemble of feature-based univariate time series classifiers to perform very well in classifying between strong and weak SEP events. Specifically, we found the supervised time series forest (STSF; Cabello et al. \citeyear{cabello2020fast}) classifier to obtain optimal results. In the present work, we implemented our earlier methodology on an extended data set and compared the performances of three models, namely summary statistic classifier (SSC), one-nearest neighbor (1NN) and STSF. Furthermore, we consider a fixed observational window of 11 hours and use multiple prediction windows of prediction interval: 05-, 15-, 30-, 45-, and 60-min to leverage on high confidence for relatively low lead times.

In our model analysis phase, we used the Youden Index (J) to understand the trade-off between the true positive rate (TPR) and the false positive rate (FPR) at different classification thresholds. That is, we estimate an optimal threshold that provides a TPR with an acceptable FPR to make decisions using our classifiers. We considered the validation set to compare the performances of the models based on multiple skill scores, such as TSS, HSS, GSS and MCC. This resulted in STSF performing better compared to SSC and 1NN at all prediction windows. Furthermore, we use an unseen test set to obtain the classification report of the best model, including the contingency tables. For a 60-minute prediction window, we obtain the following scores for STSF: TSS = 0.850; HSS = 0.878; GSS = 0.783; MCC = 0.879. In summary, our results are promising and provide a gateway for our model architecture to be used in an operational context to offer short-term predictions on the arrival of strong SEP events up to 60 minutes. The data set and coding methodology of our model implementation have been made publicly available on Harvard Dataverse: \dataset[10.7910/DVN/MWY6H7]{https://doi.org/10.7910/DVN/MWY6H7} under a Creative Commons license.

The applications we foresee for our predictive system to be useful to space agencies across the globe are two-fold. First, the system can serve as a convenient tool for forecasters, for example, at NASA's Space Radiation Analysis Group (SRAG) and NOAA-SWPC in assisting them in broadcasting highly accurate last-minute warnings to astronauts on the surface of the Moon or during extravehicular activities when outside the Earth’s magnetosphere. Second, for future interplanetary missions, such as the missions to Mars that NASA is planning, the spacecraft will spend most of its time outside of the Sun-Earth line. Therefore, SEP predictions that apply to the Earth-Moon system are of limited value when the spacecraft-Sun makes a large angle with the Earth-Sun direction. A crewed spacecraft on a journey to Mars would only have to carry a scaled-down version of the GOES instruments to produce short-term warnings of intense solar particle storms. Hence, we expect that our system can be a valuable addition to the tools that space agencies will need to protect astronauts on upcoming lunar and Mars missions.

\begin{acknowledgments}
We acknowledge the use of X-ray and proton flux data from the GOES missions made available by NOAA. We thank the anonymous reviewer for constructive comments on the manuscript that have improved the contents of the paper. Authors Petrus Martens and Berkay Aydin's contribution to this work is supported by NASA SWR2O2R Grant 80NSSC22K0272. Author Sumanth Rotti carried out this work while supported by NASA's FINESST Grant 80NSSC21K1388 and SMD grant 24-SMDSS24-0045.
\end{acknowledgments}


\software{pandas (McKinney et al. \citeyear{mckinney2010data}), numpy (Van Der Walt et al. \citeyear{van2011numpy}; Harris et al. \citeyear{harris2020array}), sklearn (Pedregosa et al. \citeyear{scikit-learn}), sktime (L{\"o}ning et al. \citeyear{loning2019sktime}; L{\"o}ning et al. \citeyear{markussktime}), matplotlib (Hunter \citeyear{hunter2007matplotlib}).}

\appendix
\section{Contingency Tables}\label{sec:CT}

The contingency table elements for SSC and 1NN on the validation set are provided in Tables \ref{tab:valssccm} and \ref{tab:val1nncm}, respectively, for each of our experiments on changing prediction windows. The values from these tables are used to estimate the skill scores that are defined in Section \ref{sec:metrics} and shown in Figure \ref{fig:ss_comp}.

\movetabledown=3mm
\begin{longrotatetable}
\begin{deluxetable*}{ll|c|c|c|c|c|c|c|c|c|c|c|c|c|}
\tablenum{4}
\tablecaption{Contingency tables for the SSC model on the validation set. \label{tab:valssccm}}
\tablewidth{0pt}
\tablehead{
\textbf{} & \textbf{} & \multicolumn{2}{c|}{\textbf{T$_5$}} & \multicolumn{2}{c|}{\textbf{T$_{15}$}} & \multicolumn{2}{c|}{\textbf{T$_{30}$}} & \multicolumn{2}{c|}{\textbf{T$_{45}$}} & \multicolumn{2}{c|}{\textbf{T$_{60}$}} \\
\textbf{} & \textbf{} & \multicolumn{2}{c|}{Predicted} & \multicolumn{2}{c|}{Predicted} & \multicolumn{2}{c|}{Predicted} & \multicolumn{2}{c|}{Predicted} & \multicolumn{2}{c|}{Predicted} \\
 &  & Strong & Weak & Strong & Weak & Strong & Weak & Strong & Weak & Strong & Weak}
\startdata
 & Strong & 56 & 24 & 43 & 18 & 42 & 38 & 44 & 36 & 58 & 22 \\
True & ||| & ||| & ||| & ||| & ||| & ||| & ||| & ||| & ||| & ||| & ||| \\
 & Weak & 44 & 850 & 37 & 876 & 15 & 879 & 20 & 874 & 45 & 849 \\
\enddata
\tablecomments{Truth tables for the SSC model for different prediction windows. The first column is a shared entry of true labels against predictive labels for each experiment. The elements indicate the number of predictions with respect to the actual occurrences in the validation set. 
\newline
Prediction windows are shown by a subscript to T. For example, T$_5$ indicates a prediction window of 5 min.
\newline
Model name: SSC - Summary Statistic Classifier}
\end{deluxetable*}
\end{longrotatetable}

\movetabledown=3mm
\begin{longrotatetable}
\begin{deluxetable*}{ll|c|c|c|c|c|c|c|c|c|c|c|c|c|}
\tablenum{5}
\tablecaption{Contingency tables for the 1NN model on the validation set. \label{tab:val1nncm}}
\tablewidth{0pt}
\tablehead{
\textbf{} & \textbf{} & \multicolumn{2}{c|}{\textbf{T$_5$}} & \multicolumn{2}{c|}{\textbf{T$_{15}$}} & \multicolumn{2}{c|}{\textbf{T$_{30}$}} & \multicolumn{2}{c|}{\textbf{T$_{45}$}} & \multicolumn{2}{c|}{\textbf{T$_{60}$}} \\
\textbf{} & \textbf{} & \multicolumn{2}{c|}{Predicted} & \multicolumn{2}{c|}{Predicted} & \multicolumn{2}{c|}{Predicted} & \multicolumn{2}{c|}{Predicted} & \multicolumn{2}{c|}{Predicted} \\
 &  & Strong & Weak & Strong & Weak & Strong & Weak & Strong & Weak & Strong & Weak}
\startdata
 & Strong & 44 & 36 & 44 & 36 & 43 & 37 & 43 & 37 & 43 & 37 \\
True & ||| & ||| & ||| & ||| & ||| & ||| & ||| & ||| & ||| & ||| & ||| \\
 & Weak & 25 & 869 & 24 & 870 & 25 & 869 & 25 & 869 & 29 & 865 \\
\enddata
\tablecomments{Truth tables for the 1NN model for different prediction windows. The first column is a shared entry of true labels against predictive labels for each experiment. The elements indicate the number of predictions with respect to the actual occurrences in the validation set. 
\newline
Prediction windows are shown by a subscript to T. For example, T$_5$ indicates a prediction window of 5 min.
\newline
Model name: 1NN - One-nearest Neighbor}
\end{deluxetable*}
\end{longrotatetable}

\bibliography{sample631}{}
\bibliographystyle{aasjournal}


%% Include this line if you are using the \edit1, \replaced, \deleted
\listofchanges
\end{document}